\documentclass[twocolumn,superscriptaddress,showpacs,prd,aps,amsmath,amssymb]{revtex4-1}
\usepackage{graphicx}
\usepackage{bm}
\usepackage{color}
\usepackage{soul}
\usepackage{hyperref}
\usepackage{enumerate}

\newcommand{\beq}{\begin{equation}}
\newcommand{\eeq}{\end{equation}}
\newcommand{\beqn}{\begin{eqnarray}}
\newcommand{\eeqn}{\end{eqnarray}}

\newcommand{\mnras}{Mon.\ Not.\ R.\ Astro.\ Soc.}

\newcommand{\maxtov}{M_{\rm max}^{\rm sph}}
\newcommand{\maxsup}{M_{\rm max}^{\rm sup}}

\begin{document}
\title{GW170817, General Relativistic Magnetohydrodynamic Simulations,
  and the Neutron Star Maximum Mass}
\author{Milton Ruiz}
\affiliation{Department of Physics, University of Illinois at
  Urbana-Champaign, Urbana, IL 61801}
\author{Stuart L. Shapiro}
\affiliation{Department of Physics, University of Illinois at
  Urbana-Champaign, Urbana, IL 61801}
\affiliation{Department of Astronomy \& NCSA, University of
  Illinois at Urbana-Champaign, Urbana, IL 61801}
\author{Antonios Tsokaros}
\affiliation{Department of Physics, University of Illinois at
  Urbana-Champaign, Urbana, IL 61801}    

\begin{abstract}
Recent numerical simulations in general relativistic magnetohydrodynamics (GRMHD)
provide useful constraints for the interpretation of the GW170817
discovery. Combining the observed data with these simulations leads to a
bound on the maximum mass of a cold, spherical neutron star (the TOV limit):
$\maxtov\lesssim 2.74/\beta$, where $\beta$ is the ratio of the maximum mass of
a uniformly rotating neutron star (the supramassive limit) over the maximum mass
of a nonrotating star. Causality arguments allow  $\beta$ to be as high as
$1.27$, while most realistic candidate equations of state predict $\beta$
to be closer to $1.2$, yielding $\maxtov$ in the range $2.16-2.28 M_\odot$.
A minimal set of assumptions based on these simulations
distinguishes this analysis from previous ones, but leads
to a similar estimate. There are caveats, however, and they are 
enumerated and discussed. The caveats can be removed by further
simulations and analysis to firm up the basic argument.
\end{abstract}
\maketitle

\section{Introduction}
The long-sought premise of multimessenger astronomy was recently
realized with the  detection of a gravitational wave (GW)
signal from a low-mass binary system by the LIGO/VIRGO detectors
\cite{TheLIGOScientific:2017qsa}. Event GW170817, which was
accompanied by a short $\gamma$-ray burst (sGRB), revealed that
if the compact objects have a low dimensionless spin ($\chi\leqslant
|0.05|$), then the inferred masses of each component of the binary
and its total mass are $m_1\in (1.36,1.60)\,M_\odot$,  $m_2\in
(1.17,1.36)\,M_\odot$, and $m_1+m_2=2.74^{+0.04}_{-0.02}\, M_\odot$,
respectively. This strongly suggests a merging binary neutron star
system (NSNS) as the source of GW170817, although it cannot rule
out the possibility that one of the binary companions is a stellar-mass
black hole (BH). Evidence that such low-mass black holes (LMBHs)
exist is very weak (see e.g.~\cite{Yang:2017gfb} for a summary of
possible LMBH formation mechanisms and routes by which they may
arise in binaries with NS companions). Since the usual mechanisms
believed to generate stellar-mass BHs, such as the collapse of massive
stars, result in BHs with masses significantly larger, we tend to
rule out a BHNS merger as a possible source of GW170817.

The coincident sGRB (GRB 170817A) of duration $T_{90}= 2\pm 0.5\,\rm s$
was detected by the Fermi Gamma-Ray Burst Monitor~\cite{2017GCN.21520....1V,
  2017GCN.21517....1K} and INTEGRAL~\cite{Savchenko:2017ffs,Savchenko17GCN}
$1.734\pm\,0.054\,$s after  the GW170818 inferred binary coalescence time,
at a luminosity distance of $40^{+8}_{-8}\,\rm Mpc$ in the galaxy
NGC 4993. Here $T_{90}$ denotes the time during which $90\%$ of the total
counts of $\gamma$-rays have been detected. The  burst exhibited
an atypically low luminosity ($L\sim 10^{47}\,\rm erg/s$)
and the absence of an afterglow during the first days, which has been attributed
to the off-axis viewing of GRB emission (see e.g.~\cite{Shibata:2017xdx,
  Fraija:2017aev}). It is likely that its volumetric value is much
larger and comparable to typical sGRB values. Subsequent optical/infrared
transients consistent with kilonova/macronova models were also observed
(see e.g.~\cite{GBM:2017lvd,Monitor:2017mdv, Abbott:2017wuw}).

One of the most important puzzles in high energy astrophysics is the
ground state of matter at zero temperature, which is closely related to the
maximum gravitational mass, $\maxtov$, of a nonrotating, spherical NS
\footnote{{GW170817 is also consistent with the coalescence of a binary 
hybrid hadron-quark star-neutron star \cite{2017arXiv171200451P}.}  }.
To date the largest pulsar masses observed are $2.01\pm0.04 M_\odot$ for
J0348+0432 \cite{2013Sci...340..448A}, and $1.928\pm0.017 M_\odot$ for
J1614-2230 \cite{2010Natur.467.1081D}, but the quest for a firm upper limit
on the mass of a NS has a long history \cite{1978PhR....46..201H}
that started in 1974 by Rhoades and Ruffini \cite{1974PhRvL..32..324R}.
Their argument involved a matching mass-energy density $\rho_m$ below which
the equation of state (EOS) is well known, while from that point on a causal
EOS for the pressure $P$ ($P=\rho+\textrm{const}$) is invoked.\footnote{Although the radius
  of the star is  controlled by the low density part of the EOS the
  maximum mass that can be supported is controlled by the high density
  region~\cite{2016PhR...621..127L}.}
This upper mass limit depends on the matching density \cite{1977ApJ...213..831H},
and assuming $\rho_m=4.6\times 10^{14}\,\rm {gr/cm^3}\approx 1.7
\rho_{\rm nuc}$ they obtained an upper limit of $\maxtov=3.2M_{\odot}$.
As the matching density increases the maximum mass for a spherical star
decreases as $\rho_m^{-1/2}$. 
For example, in~\cite{1996ApJ...470L..61K} where the confidence
of the EOS was taken to be up to  $\rho_m=2 \rho_{\rm nuc}$, a
maximum mass of $2.9\ M_\odot$ was obtained (see \cite{2017arXiv170704966B}
for recent review). In~\cite{Read:2008iy} a parametrized piecewise-polytropic
fitting was introduced in order to make a systematic study of different
constraints placed on  high density, cold matter, including the causality
constraint. More recently and from another point of view, based on the sGRB
scenario, a survey of a wide EOS parameter space and matching  densities using
plausible masses for a NS merger remnant concluded that $\maxtov\approx 2-2.2M_\odot$
\cite{2015ApJ...808..186L}. At high matching densities the core has negligible
mass and the upper bound becomes independent of $\rho_m$. In
\cite{2015ApJ...812...24F} Newtonian merger simulations with different EOS,
resulted in an upper bound of $2.4M_\odot$.

With the discovery of GW170817, \cite{2017arXiv171005938M} used
electromagnetic (EM) constraints on the remnant imposed by the
kilonova observations after the merger, together with GW information, to make
a tight prediction of~$\maxtov\leq 2.17M_\odot$ with $90\%$ confidence. They
argued that the NSNS merger resulted in a \textit{hypermassive} NS (HMNS;
\cite{BaShSh}) that collapses
to a BH in $\approx 10-1000{\rm ms}$, producing the observed kilonova ejecta
expanding at mildly relativistic velocities. By contrast,
\cite{2017arXiv171007579S}  summarized a broad number of their
relativistic hydrodynamic simulations favoring a long-lived, massive
NS surrounded by a torus to support their inferred requirement of a strong neutrino
emitter that has a sufficiently high electron fraction to avoid an enhancement
of the ejecta opacity. To get such remnants one needs an EOS with a high value
of $\maxtov$, which they place between $2.15-2.25M_\odot$. Although the authors
disfavor the scenario of a BH-accretion disk system as a weak neutrino emitter,
they remind us that there is no current consensus and more simulations are
needed to address this point.

Here we  introduce a new ingredient into the arguments put forward previously
to narrow down  possible merger outcomes. In particular we focus on GRMHD
simulations that we performed recently, as well as some that we are currently
performing, 
to argue that to have a sGRB as in GW170817 the merger remnant is likely
an HMNS that undergoes delayed collapse. We also pinpoint
how the GW170817 data can be combined with causality arguments to
establish an interesting NS upper mass limit. In the process we identify the
caveats that underlie this determination and thereby  indicate areas for future
investigation.

\section{Assumptions}
\label{sec:mhdsim}

In this work we make the following assumptions which we justify in the
following section.

\begin{enumerate}[(i)]
\item GW170817 and the associated sGRB result from the merger of an NSNS;
\item A BH arises following the formation of an
      HMNS that undergoes delayed collapse soon after the NSNS merger;
\item The sGRB is powered by the  spinning BH accreting
      gas from a circumstellar, magnetized disk formed from NS tidal debris;
    \item To trigger the sGRB a collimated, magnetically confined, helical jet was
      launched from the poles of the spinning BH remnant.
\end{enumerate}

\section{Justification and Caveats}
\label{sec:mhdsim}
Here we discuss the above assumptions in order.

\begin{enumerate}[(i)]
  
\item NSNS and BHNS mergers are the most promising and widely
  accepted models for a central engine capable of powering an
  sGRB~\cite{Blinnikov84,EiLiPiSc,NaPaPi,
  Pacz86,Piran:2002kw,bergeretal05,Foxetal05,hjorthetal05,
  bloometal06,Baiotti:2016qnr,Paschalidis:2016agf}, which
  is supported by the {\it first} detection of a
  kilonova associated with the sGRB ``GRB 130603B''
  \cite{Tanvir:2013pia,Berger:2013wna}. 
  Given little evidence for LMBHs in the mass range compatible
  with either companion in GW170817 ($\lesssim 1.6M_{\odot}$), we consider only
  NSNS systems in our analysis.

\item All GR simulations show that immediately after the NSNS merger
  the remnant is a differentially rotating NS that has undergone shock
  heating following contact. The configuration either settles into
  quasiequilibrium in a couple of rotation periods, or undergoes prompt
  collapse.
  A combination of GW emission, magnetic winding,
  turbulent viscosity and neutrino cooling radiate away or
  redistribute some of its angular momentum, driving the core toward
  rigid rotation~\cite{Duez:2004nf}. There are several possible final
  outcomes that can be split into two broad categories
  \cite{Baumgarte10,2013rehy.book.....R}:
\begin{enumerate}[1.]
\item  The NS remnant has a mass below the \textit{supramassive} limit
  \cite{CookShapTeuk}, which for a given EOS is the maximum mass of a uniformly
  rotating NS. This case can lead to the following possible scenarios:
  \begin{enumerate}
  \item[1a.] If the mass is smaller than the maximum spherical
    limit, $\maxtov$, then it will live for a very long-time as a spinning
    NS until, e.g., pulsar magnetic dipole emission causes its spindown
    once the tidal debris is cleared away. Without a BH
    and its ergosphere~\cite{Ruiz:2012te} to launch  a collimated jet and
    power an sGRB  after merger, this scenario is unlikely to have occurred
    in GW170817.

  \item[1b.] If the mass of the remnant is larger than $\maxtov$ then the same
    mechanism as in (1a) will lead to BH formation close to the turning 
    point \cite{1988ApJ...325..722F,2011MNRAS.416L...1T}. 
    This possibility is 
    also unlikely since the spindown timescale to BH formation is longer
    than the observed times from GW170817 ($\sim 100$s~\cite{ShaTeu83a}).
    [But see~\cite{Shibata:2017xdx} for an alternative argument based on kilonova
    considerations.]
  \end{enumerate}

\item The remnant has a mass above the supramassive limit, i.e.,
      it is a (transient) HMNS supported primarily by 
      differential rotation \cite{BaShSh}.
\end{enumerate}
      
  \begin{enumerate}
  \item[2a.] If its mass is below a critical threshold, $M_{\rm thresh}$,
        which also depends on the EOS,
        the HMNS will persist for many orbital periods, but
        eventually will undergo delayed collapse as its angular momentum
        support is driven off by GWs, magnetic winding, etc. Little mass
        is ultimately  ejected (numerical simulations indicate 
        $\lesssim 0.01 M_\odot$~\cite{Shibata:2017xdx} although 
        observations suggest that this number can be
        up to $0.05 M_\odot$ \cite{Kasliwaleaap9455}) and
        $\lesssim 5\%$ of the total rest-mass of the system goes into a
        disk around the BH~\cite{Shibata:2017xdx,Ruiz:2016rai}, depending 
        on the EoS. 
  \item[2b.] Above $M_{\rm thresh}$, the HMNS will undergo prompt collapse
        on a short dynamical timescale~\cite{STU1,Baumgarte10}.
  \end{enumerate}

Case (1b) results when the merger remnant of an NSNS is a highly magnetized,
supramassive NS~\cite{CookShapTeuk}.  We recently modeled this scenario by considering
a differentially rotating NS seeded with an interior and exterior  dipole-like
magnetic field that is initially dynamically unimportant~\cite{ruiz2017b}.
This case is depicted in Fig. 1, first column.
Our preliminary numerical simulations do not exhibit evidence of jet formation, nor
does  a force-free magnetosphere arise, which is necessary for, e.g., the
Blandford-Znajeck  (BZ) mechanism to power a collimated jet Poynting luminosity.
Thus remnant NS, which may arise and
live arbitrarily long following the merger of a NSNS, probably cannot power
an sGRB. However, very long simulations are required to completely rule out
this scenario. On the other hand, depending on the strength
of the magnetic field,  the above remnant can be a  long-lived magnetar.
Now~\cite{2017arXiv171005938M} points out that delayed X-ray emission observed
after many sGRBs suggest the presence of magnetars  and 
raise doubt about whether BH formation is a strict requirement to produce sGRBs.
But GW170817 showed
no evidence for such high-energy emission~\cite{Monitor:2017mdv}, which
is a property of a long-lived magnetar~\cite{Metzger:2007cd,Rowlinson:2013ue}.
A  magnetar remnant can release part of its rotational energy
through strong magnetic dipole radiation, which accelerates the ejected matter
and produces very bright X-ray emissions~\cite{Shibata:2017xdx}. These two features
are not observed along with the optical/infrared transients in GW170817~\cite{Monitor:2017mdv,
  GBM:2017lvd}, and therefore, the magnetar scenario is also unlikely \cite{2017arXiv171005938M}.
Isolated NSs can, in principle, trigger the BZ  mechanism if the NS
spacetime exhibits ergoregions~\cite{Komissarov:2004ms,Ruiz:2012te}. However,
NSs supported by realistic EOSs probably cannot form ergoregions~\cite{1978MNRAS.182...69S}.
Since these configurations are unstable or marginally stable under scalar
and EM perturbations~\cite{friedman1978} the probability of their significance 
is low. 

Case (2a) is the scenario most favored by our GRMHD simulations and is depicted in Fig. 1,
second column. It gives rise
to an interesting upper mass limit which we discuss in Section \ref{sec:ml}.

For scenario (2b) our recent GRMHD simulations ~\cite{Ruiz:2017inq}
show that remnants that undergo
prompt collapse do not have time to sufficiently amplify their  poloidal magnetic fields
and overcome the ram pressure of the residual tidal debris near the
polar regions of the star. As a result
the polar regions above the resulting BH never achieve the necessary force-free conditions
(characterized by $B^2/(8\pi\rho_0) \gg 1$, as in a pulsar magnetosphere, 
where $B$ is the magnetic field and $\rho_0$ is
the rest mass density) capable of launching a jet and powering an sGRB. This case is
depicted in Fig. 1, third column.
Therefore these preliminary simulations tend to rule out the prompt collapse scenario as a
possible origin of GW170817.

\item A spinning BH-disk system is the most promising and widely accepted model for a
central engine capable of powering an sGRB.
Several distinct mechanisms
invoking such a system have been proposed to trigger an
sGRB~\cite{EiLiPiSc,LeeRamirezRuiz2007,Meszaros:2006rc,MoHeIsMa,
  Nakar2007,NaPaPi,prs15}. One of the most successful and the subject of the most
detailed studies is the  BZ mechanism~\cite{BZeffect},
which requires the disk to be threaded by a poloidal magnetic field that
connects gas in the disk to footpoints near the BH poles. Alternative
sGRB mechanisms driven by neutrinos alone appear to be inadequate
\cite{Just:2015dba,Perego:2017fho}, so magnetic fields likely play a
crucial role.  Many numerical simulations have been
performed in GRMHD of magnetized disk accretion onto spinning BHs
in stationary Kerr spacetimes
\cite{McKinney:2006sc,2004ApJ...611..977M,dhk03,
  GRMHD_Jets_Req_Strong_Pol_fields,Hawley2001}. The accretion exhibits
BZ behavior and results in a significant Poynting jet luminosity.
Our recent GRMHD numerical  simulations of magnetized, merging
NSNSs  that form HMNSs and undergo delayed collapse track the late binary inspiral,
through merger and BH-disk formation, and then continue until a quasistationary
state is reached~\cite{Ruiz:2016rai}. The result is a spinning BH remnant in
a highly magnetized circumstellar disk that can power an sGRB via
the BZ mechanism. 

\begin{figure*}
  \centering
  \includegraphics[width=0.31\textwidth]{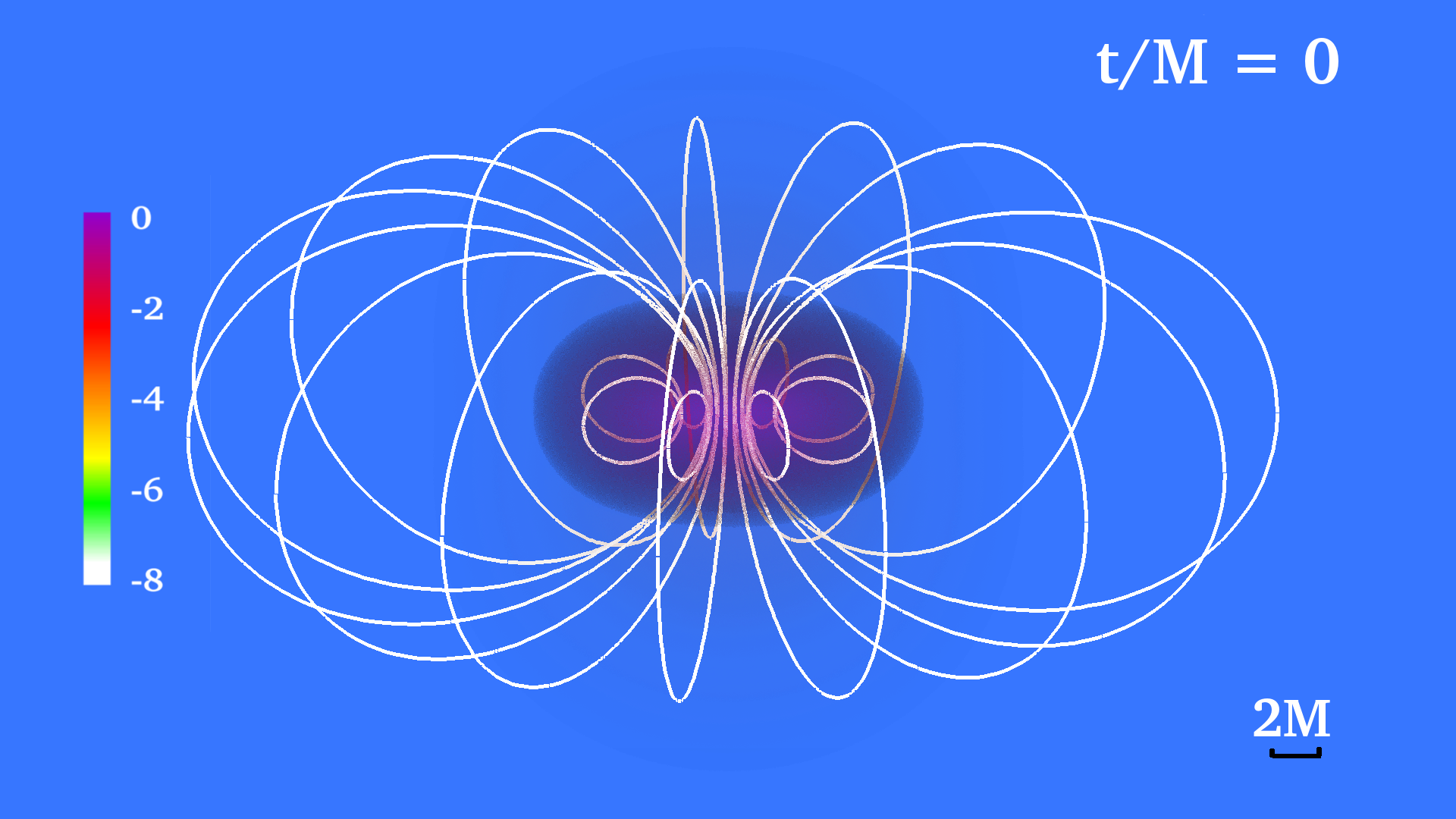}
  \includegraphics[width=0.31\textwidth]{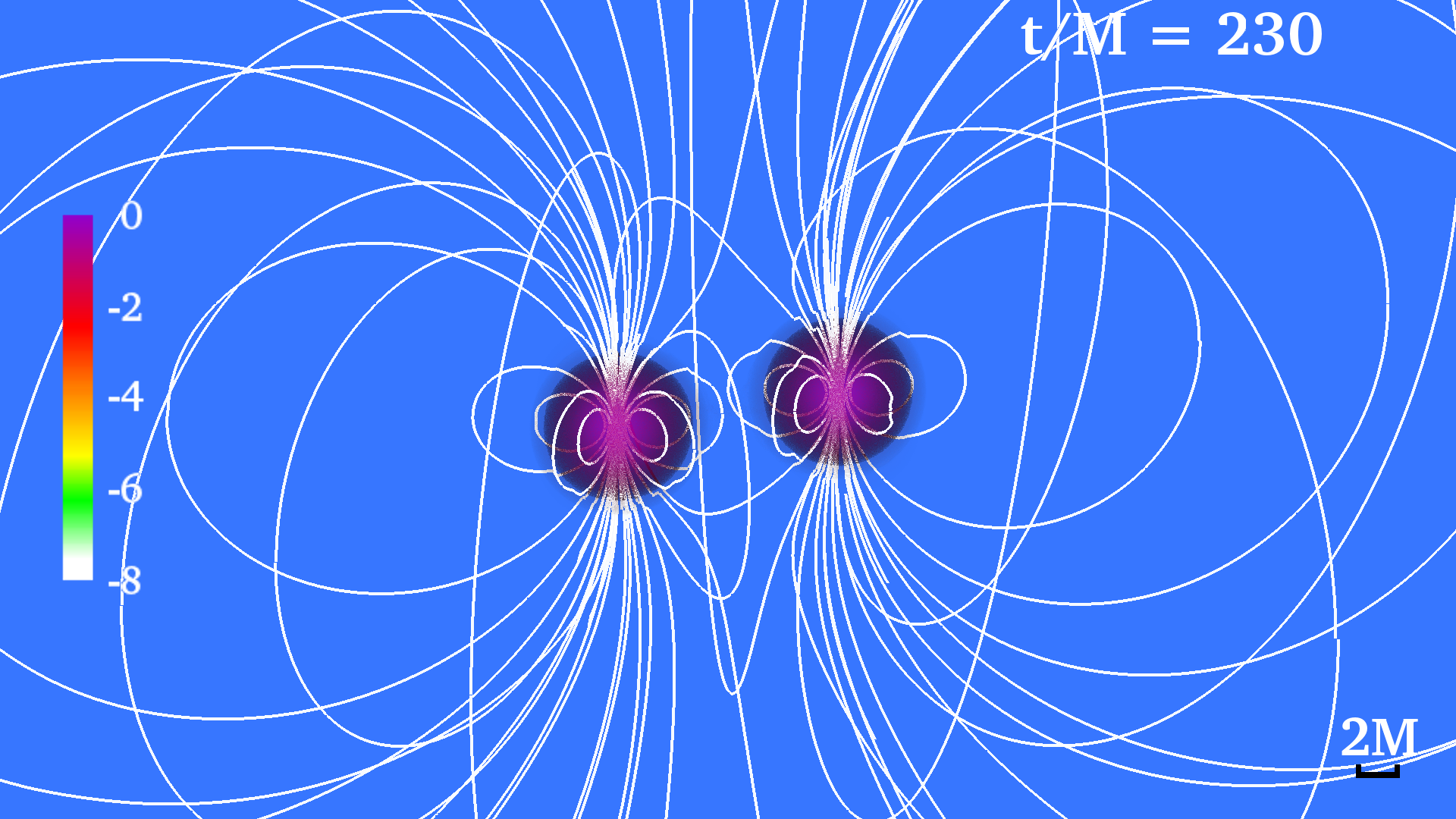}
  \includegraphics[width=0.31\textwidth]{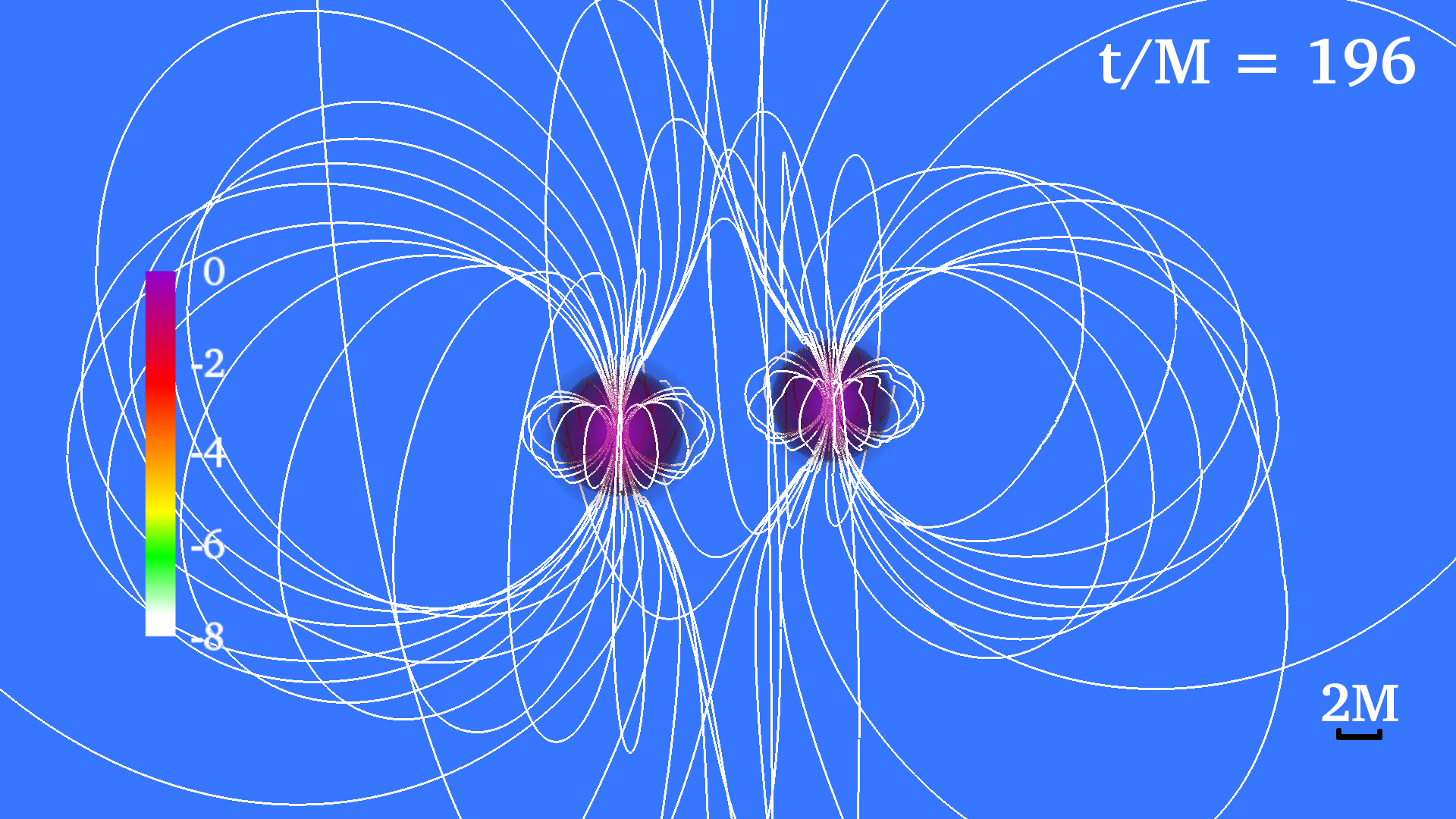}
  \includegraphics[width=0.31\textwidth]{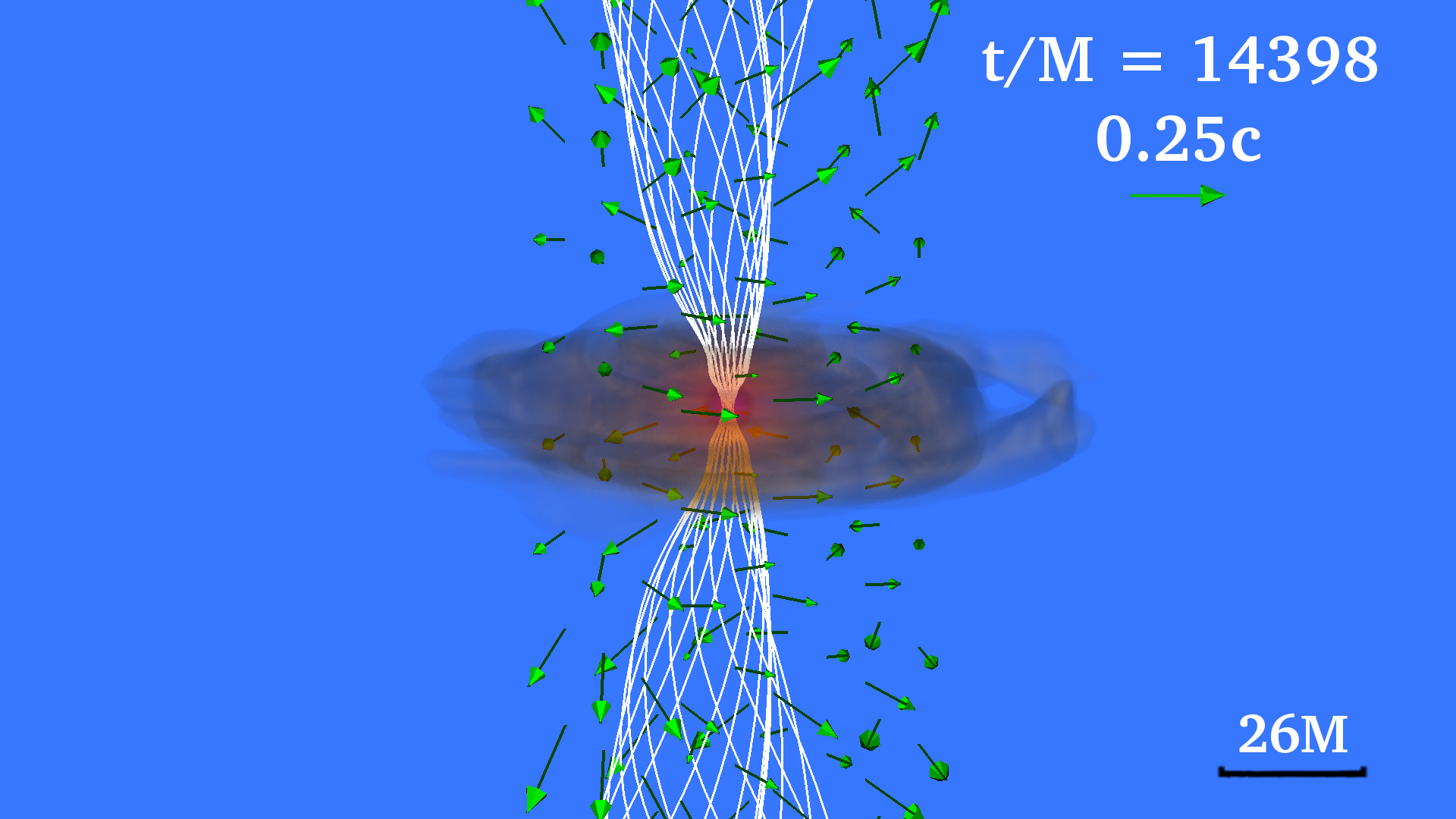}
  \includegraphics[width=0.31\textwidth]{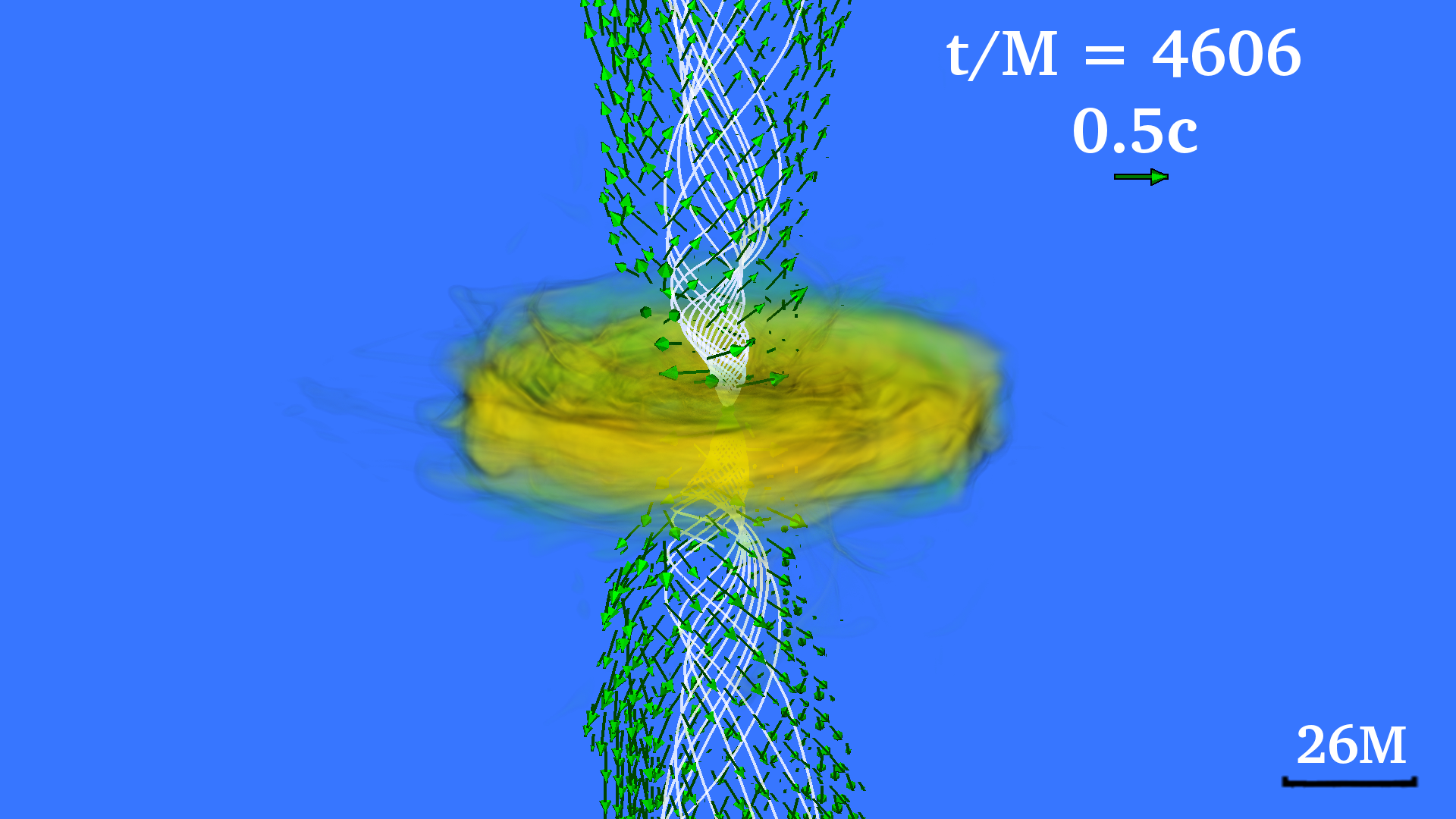}
  \includegraphics[width=0.31\textwidth]{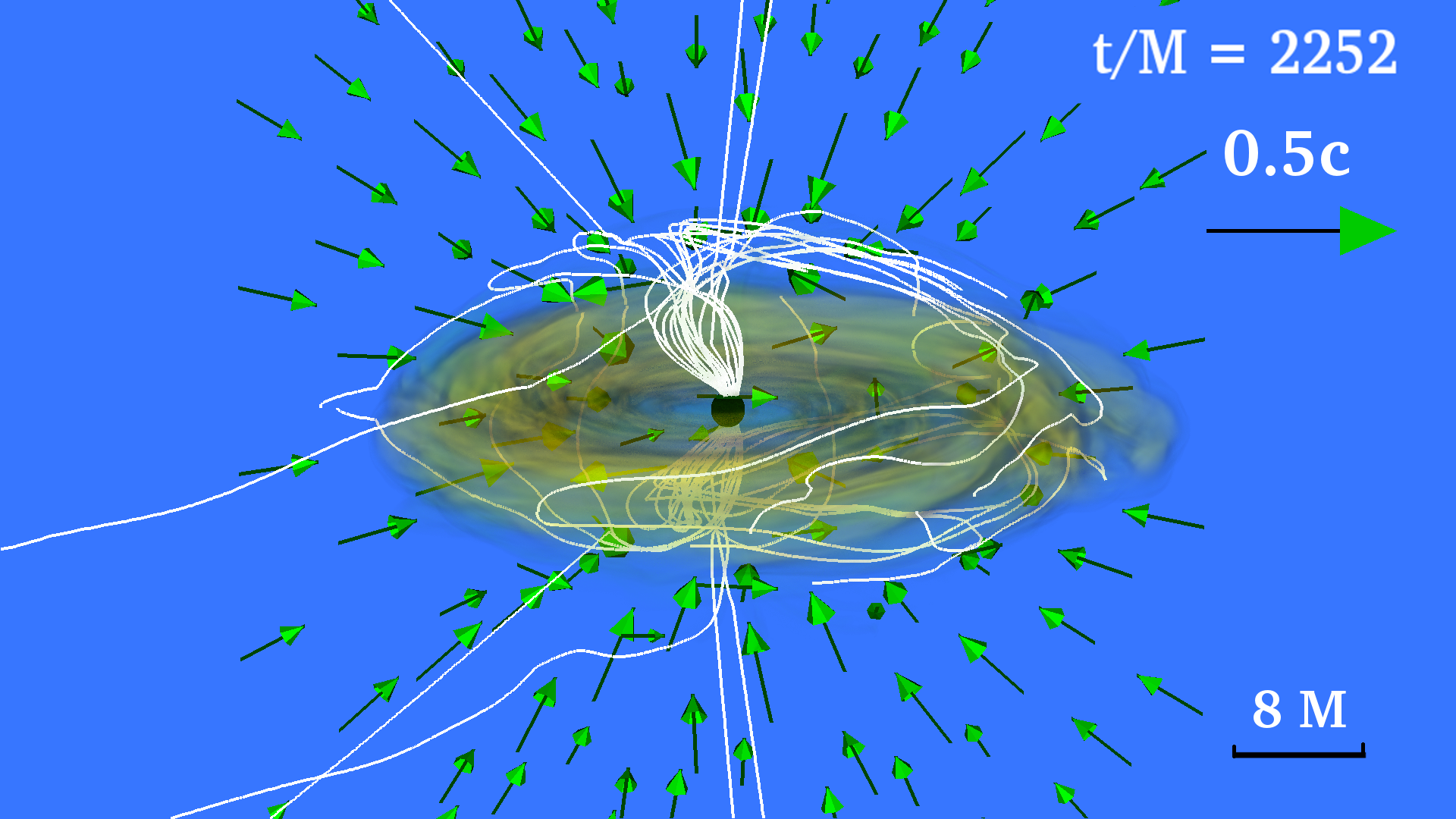}
  \includegraphics[width=0.31\textwidth]{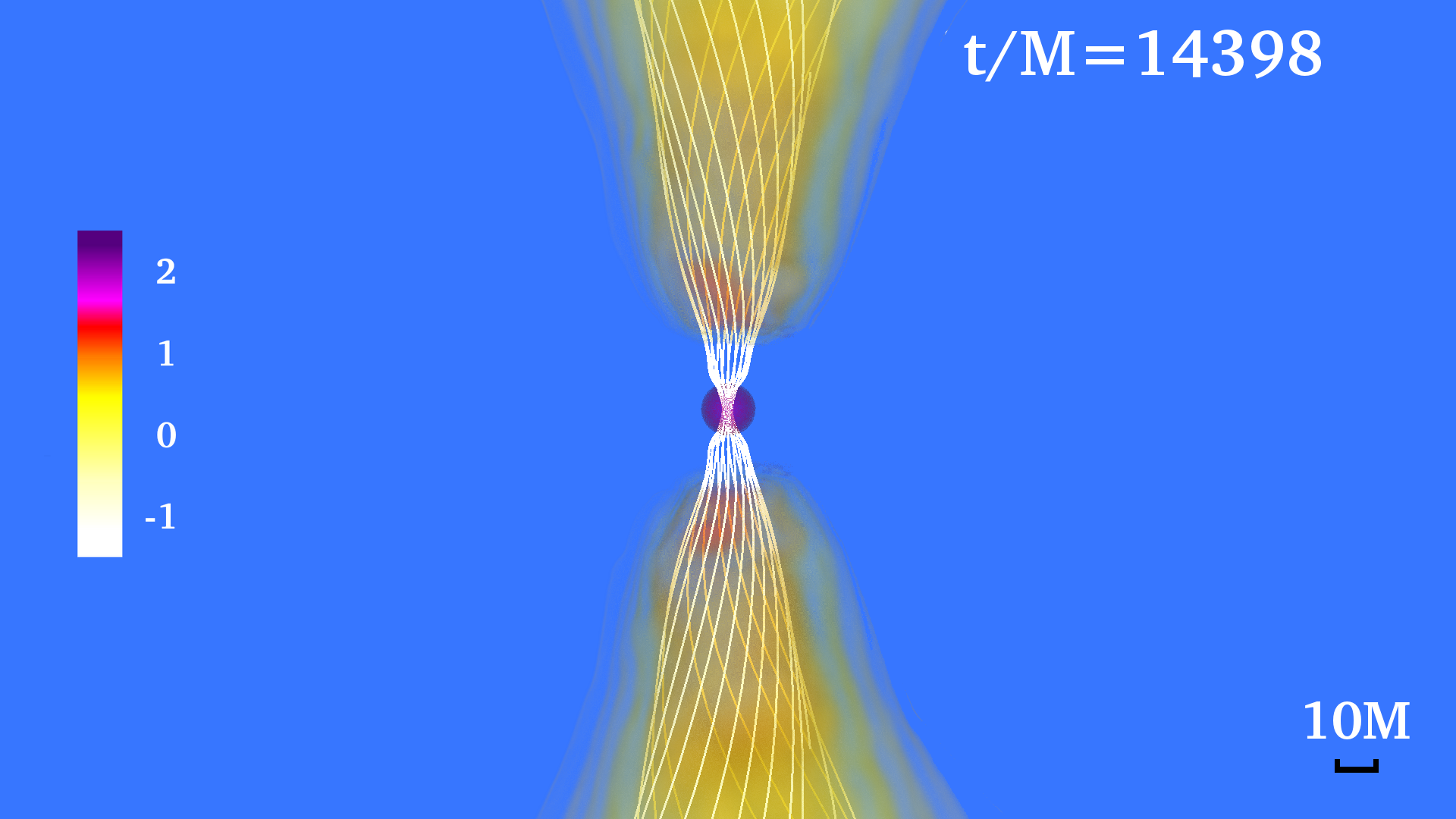}
  \includegraphics[width=0.31\textwidth]{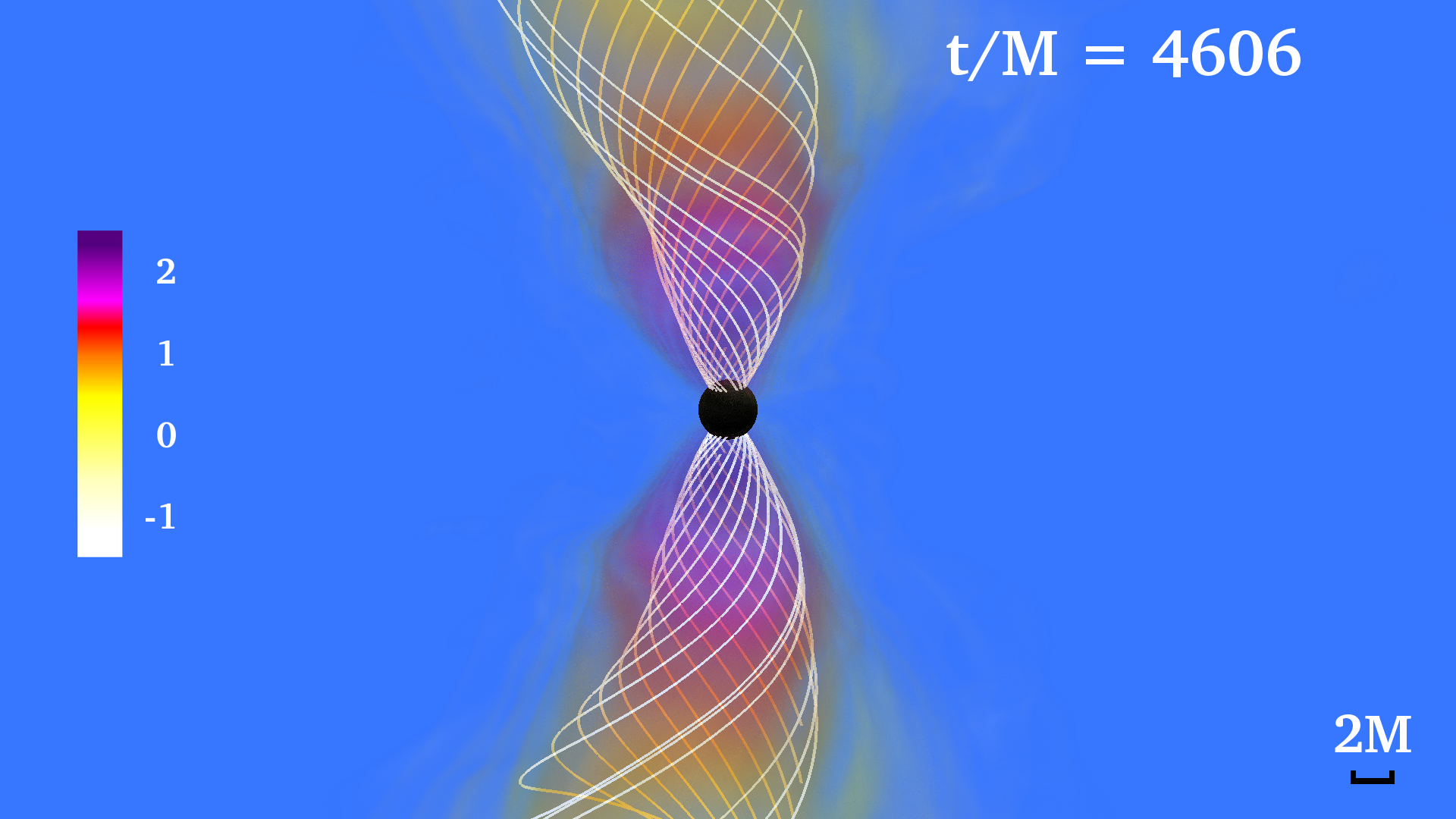}
  \includegraphics[width=0.31\textwidth]{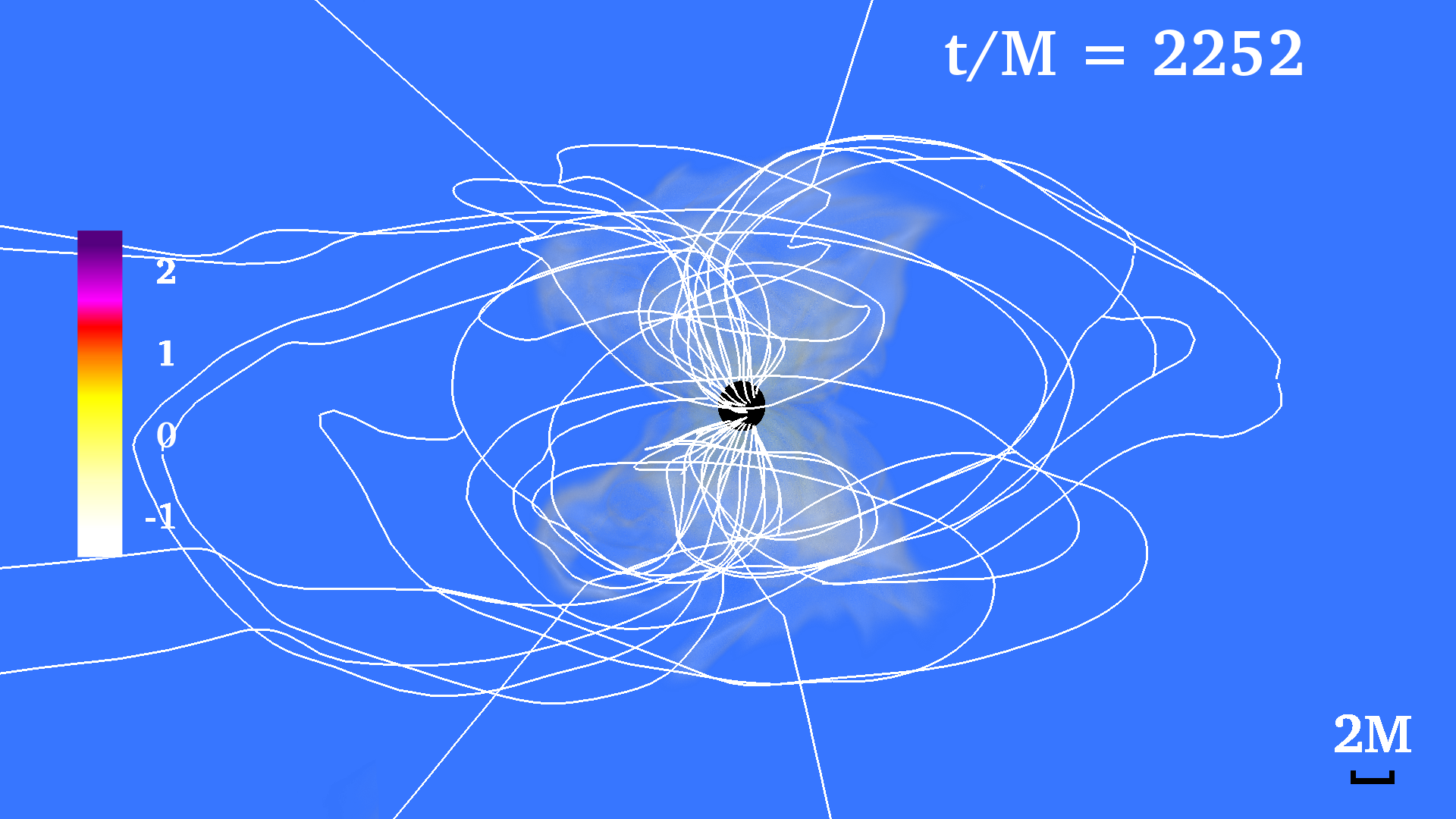}
  \caption{
Snapshots of the rest-mass density, normalized to its initial maximum value (log scale), for
a supramassive remnant that persists, (first column, case (1b)),
a HMNS remnant that undergoes delayed collapse (second column, case (2a)), 
a HMNS remnant that undergoes prompt collapse (third column, case (2b)).
First row shows NSs at the time of B-field insertion; second
row shows the final quasistationary remnants. Third row shows the $B^2/(8\pi\rho_0)$ field
(log scale), the magnetic lines emanating from the remnant poles, and the fluid velocity flow vectors.
The field lines form a tightly wound helical funnel and drive a jet in delayed collapse,
but not in the other two cases. Here $M=0.0136(M_{\rm tot}/2.74 M_\odot){\rm ms} =4.07(M_{\rm tot}/2.74 M_\odot)$km.
\label{fig:rmd}}
\end{figure*}

\item Most models of sGRB require that the main conduit through which energy
flows from the central BH engine to the outer regions where $\gamma$-rays
and other forms of EM emission are generated is a relativistic jet. Such
a jet is a key signature of the BZ mechanism.  The jet is confined
inside a force-free region  by a tightly wound,
helical magnetic field that emanates from the poles of the spinning BH and
extends to very large radii. Our preliminary simulations show the clear formation
of a bonafide incipient jet with these characteristics~\cite{Ruiz:2016rai}
(see Fig.~\ref{fig:rmd}, column 2).
In particular, we found that NSNS mergers that undergo delayed collapse
can launch a magnetically-driven jet after $\sim 43(M_{\rm NS}/1.625M_\odot)$ms
following the merger. 
The disk lifetime, which may be comparable to the burst duration, $T_0$, 
and its EM luminosity, were found to be $\tau_{\rm disk}\sim 0.1$s and
$L_{EM}\sim 10^{51}\rm erg\, s^{-1}$, respectively,
consistent with typical short sGRBs (see e.g.~\cite{Kann:2008zg, Shapiro:2017cny}).
Different EOSs, NS mass ratios and spins, initial B-field topologies, and possibly neutrinos, 
affect the amount and composition of the ejecta
during NSNS coalescences~\cite{Hotokezaka2013,Palenzuela:2015dqa,Ciolfi:2017uak,
Sekiguchi2015,Foucart:2016rxm,Lehner:2016lxy}, and  therefore, the
ram pressure produced by the fall-back debris, as well as the mass
of the accretion disk. The delay time for jet launching following the
merger, as well as its lifetime, may therefore depend on these parameters. This may
explain the delay time ($\sim 1.7$s) between the inferred merger time of GW170817 and 
the sGRB. Other processes associated with the sGRB $\gamma$-ray energy reprocessing
mechanism may also contribute to the delay.
\end{enumerate}
A few caveats remain regarding the GRMHD simulations in~\cite{Ruiz:2017inq} and
\cite{Ruiz:2016rai}.
These preliminary equal-mass, irrotational binary studies adopt idealized, $\Gamma$-law
EOSs ($\Gamma=2$), adiabatic evolution (except for shocks), and aligned initial magnetic
fields. They enabled us to provide a ``proof of principle'' regarding jet formation
and Poynting jet luminosity beams. That GW170817 has an atypically low $\gamma$-ray
luminosity ($L\sim 10^{47}\,\rm erg/s$) several orders of magnitude smaller than typical
sGRBs, can be attributed to an off-axis viewing \cite{Shibata:2017xdx,Fraija:2017aev}.
A more extensive parameter survey incorporating realistic hot, nuclear EOSs, 
different binary mass ratios and spins, different initial B-field topologies, and neutrino
transport are required to corroborate the above conclusions. 

\section{Maximum mass}
\label{sec:ml}

Invoking scenario (2a) of Section \ref{sec:mhdsim}, favored by our GRMHD simulations,
and noting that  the total initial gravitational mass of the NSNS binary in GW170817 is
$M_{\rm NSNS} \approx 2.74 M_{\odot}$ we have
\begin{equation}
  \label{eq1}
  M_{\rm NSNS} \approx 2.74 \lesssim M_{\rm thresh} \approx \alpha \maxtov.
\end{equation}
Here $\alpha \approx 1.3 - 1.7$ is the ratio of the 
HMNS threshold mass limit to the NS spherical maximum mass as gleaned from multiple
numerical experiments of merging NSNSs 
\cite{Shi05,Shibata:2006nm,Baiotti:2008ra,Hotokezaka:2011dh,Bauswein:2013jpa}. 
We also have that
\begin{equation}
  \label{eq2}
  M_{\rm NSNS} \approx 2.74 \gtrsim \maxsup \approx \beta \maxtov,
\end{equation}
where $\beta \approx 1.2$ is the ratio of the uniformly rotating
supramassive NS limit to the nonrotating spherical maximum mass
as determined by numerical studies of multiple candidates for
realistic cold, nuclear EOSs \cite{Cook:1993qj, Cook:1993qr, Breu:2016ufb}. 
Now if one uses a Rhoades-Ruffini causality argument for the maximum mass of
a nonrotating as well as a uniformly rotating star 
\cite{1987ApJ...314..594F,1997ApJ...488..799K}, one obtains
\beqn
\maxtov\ & =\ & 4.8 \left(\frac{2\times10^{14}\ {\rm gr/cm^3}}{\rho_m/c^2}\right)^{1/2} M_\odot\, ,  \\
\maxsup\ & =\ & 6.1 \left(\frac{2\times10^{14}\ {\rm gr/cm^3}}{\rho_m/c^2}\right)^{1/2} M_\odot\, ,  
\eeqn
from which $\beta \approx 1.27$.
Combining Eqs. (\ref{eq1}), (\ref{eq2}) we arrive at
\begin{equation}
2.74/\alpha \lesssim \maxtov \lesssim 2.74/\beta
\label{eq3}
\end{equation}
Since $\alpha \gtrsim \beta > 1$, Eq.~(\ref{eq3}) is
self-consistent. Adopting the causal EOS above sets a TOV mass limit
as low as $\maxtov \lesssim 2.16$. However, most existing realistic EOS
candidates give $\beta=1.2$ (see \cite{Cook:1993qr,Lasota1996,Breu:2016ufb},
which gives a more conservative mass limit $\maxtov \lesssim 2.28$.

Uncertainties in the EOS allow the high value of $\beta$ that yields the  
low maximum mass. In particular, an EOS might be sufficiently stiff
above a some (matching) density to produce a stiff core 
and still be compatible with current NS observations, including GW170817. 
Such a core is allowed by our causal EOS and is responsible for the high value 
of $\beta$. Our inability to calculate the dense matter EOS in QCD from first 
principles permits such a possibility. On the other hand, the discovery of a 
NS with a higher mass would rule out the presence of such a stiff core.

The error bar on this upper limit comes from the uncertainty in the 
  remnant mass as well as the $\beta$ parameter. According to the LIGO/VIRGO
  measurement, the mass on the numerator in Eq. \ref{eq3}  is
$2.74^{+0.04}_{-0.01}$, i.e. it has an uncertainty of the order of $1.5\%$, 
and additionaly one has to take 
into account that a small percentage of the total mass will be ejected
or radiated away. So, we can conservatively expect a total error less than
$6.5\%$ ($1.5\%$ from the LIGO/VIRGO mass measurement plus $\sim 5\%$ from 
possible ejecta and radiation)
for the remnant mass that comes into Eq. \ref{eq3}.
The uncertainty in the $\beta$ parameter is more difficult to estimate since
there is no systematic study of causal realistic EOSs currently. For
a spread of realistic EOSs this error is close to $2\%$ \cite{Breu:2016ufb}
and the expectation is
that will remain the same when one considers causal EOS since both the 
spherical maximum and the Keplerian limit will be affected in the same way.
Combining both errors one arrives at a maximum mass $\maxtov$ between
$2.16\pm 0.23$ and $2.28\pm 0.23$. In this analysis we considered low-spin priors
$|\chi|<0.05$ since the fastest known pulsar in a binary has $\chi\leq 0.05$ and
in general one does not expect spins higher than $0.5$ \cite{Brown:2012qf}. If one
assumes instead that $\chi\leq 0.89$ for which the  remnant mass is
$2.82_{-0.09}^{+0.47}$~\cite{TheLIGOScientific:2017qsa} then the limit will be modified
to give $\maxtov$ between $2.22\pm 0.66$ and $2.35\pm 0.66$, but we regard this
as unlikely.

Corroboration of the GRMHD scenario requiring delayed collapse 
of an HMNS remnant to generate a jet and a sGRB
central engine will support this low estimate for the
maximum mass.

\medskip


\acknowledgements
It is a pleasure to thank G. Baym, C. Gammie, F. Lamb, V. Paschalidis, E. Zhou
and K. Ury\=u for useful discussions. We also thank the Illinois Relativity
group REU team, E. Connelly, J. Simone, I. Sultan and J. Zhu
for assistance in creating Fig.~1. This work has been supported in part by
National Science Foundation (NSF) Grants PHY-1602536 and PHY-1662211, and NASA
Grants NNX13AH44G and 80NSSC17K0070 at the University of Illinois at Urbana-Champaign.
This work used the Extreme Science and Engineering Discovery Environment (XSEDE),
which is supported by NSF Grant No. OCI-1053575. This research is part of
the Blue Waters sustained-petascale computing project, which is
supported by the National Science Foundation (Award No. OCI
07-25070) and the state of Illinois. Blue Waters is a joint effort of
the University of Illinois at Urbana-Champaign and its National Center
for Supercomputing Applications.
     
\bibliographystyle{apsrev4-1}        
\bibliography{references}            

\end{document}